%% file: ms.tex
\documentclass[preprint,3p]{elsarticle}

\usepackage{times}
\usepackage{bm}
\usepackage{mathtools}
\usepackage{amssymb,amsmath,amsthm, amsfonts}
\usepackage{graphicx}

\usepackage{booktabs} 

\newif\ifproofs
\proofsfalse

\usepackage[linesnumbered,vlined, boxed, ruled]{algorithm2e}
\usepackage[hyphens]{url}
\let\oldnl\nl
\newcommand{\nonl}{\renewcommand{\nl}{\let\nl\oldnl}}
\SetKwComment{Comment}{//}{}

\makeatletter
\renewcommand{\algocf@captiontext}[2]{#1\algocf@typo. \AlCapFnt{}#2} 
\def\@algocf@capt@plain{top}
\renewcommand{\algocf@makecaption}[2]{%
  \addtolength{\hsize}{\algomargin}%
  \sbox\@tempboxa{\algocf@captiontext{#1}{#2}}%
  \ifdim\wd\@tempboxa >\hsize
    \hskip .5\algomargin%
    \parbox[t]{\hsize}{\algocf@captiontext{#1}{#2}}
  \else%
    \global\@minipagefalse%
    \hbox to\hsize{\box\@tempboxa}
  \fi%
  \addtolength{\hsize}{-\algomargin}%
}
\makeatother

\DeclarePairedDelimiter\ceil{\lceil}{\rceil}
\DeclarePairedDelimiter\floor{\lfloor}{\rfloor}

\DeclareMathOperator{\mean}{E}
\DeclareMathOperator{\var}{\mathrm{var}}
\DeclareMathOperator{\frc}{\mathrm{frac}}
\DeclareMathOperator{\pr}{Pr}
\DeclareMathOperator{\X}{\mathcal{X}}
\DeclareMathOperator{\Y}{\mathcal{Y}}
\newcommand{\prob}[1]{\prb(#1)}
\newtheorem{theorem}{Theorem}

\newcommand{\yp}{{\tilde p}}

\newcommand\blfootnote[1]{%
  \begingroup
  \renewcommand\thefootnote{}\footnote{#1}%
  \addtocounter{footnote}{-1}%
  \endgroup
}

\newenvironment{romlist}{

  \begin{enumerate}
  }{\end{enumerate}}

\journal{Information Processing Letters}

\begin{document}

\begin{frontmatter}



\title{Exact PPS Sampling with Bounded Sample Size\blfootnote{Declarations of interest: none.}}

\author[inst1]{Brian Hentschel}

\affiliation[inst1]{organization={Harvard University},
            city={Cambridge},
            state={MA},
            country={U.S.A.}}

\author[inst2]{Peter J. Haas}
\author[inst3]{Yuanyuan Tian\footnote{Work done while the author was at IBM Research.}}

\affiliation[inst2]{organization={University of Massachusetts Amherst},
            city={Amherst},
            state={Massachusetts},
            country={U.S.A.}}
            
\affiliation[inst3]{organization={Microsoft Gray Systems Lab},
            city={Mountain View},
            state={California},
            country={U.S.A.}}

\begin{abstract}
Probability proportional to size (PPS) sampling schemes with a target sample size aim to produce a sample comprising a specified number $n$ of items while ensuring that each item in the population appears in the sample with a probability proportional to its specified ``weight'' (also called its ``size''). These two objectives, however, cannot always be achieved simultaneously. Existing PPS schemes prioritize control of the sample size, violating the PPS property if necessary. We provide a new PPS scheme, called EB-PPS, that allows a different trade-off: EB-PPS enforces the PPS property at all times while ensuring that the sample size never exceeds the target value $n$. The sample size is exactly equal to $n$ if possible, and otherwise has maximal expected value and minimal variance. Thus we bound the sample size, thereby avoiding storage overflows and helping to control the time required for analytics over the sample, while allowing the user complete control over the sample contents. In the context of training classifiers at scale under imbalanced loss functions, we show that such  control yields superior classifiers. The method is both simple to implement and efficient, being a one-pass streaming algorithm with an amortized processing time of $O(1)$ per item, which makes it computationally preferable even in cases where both EB-PPS and prior algorithms can ensure the PPS property and a target sample size simultaneously. 
\end{abstract}




\end{frontmatter}



\input{introduction}

\input{ExactMarginalWeightedSampling}

\input{FractionalSamples}

\input{AlgorithmRuntime}

\input{ExampleClassifier}

\input{conclusion}



\bibliographystyle{plainnat} 
\bibliography{paper-ref}





\end{document}

%% file: introduction.tex
\section{Introduction}

As increasing dataset sizes outpace growth in computer storage and processing speeds, sampling is increasingly central to data analysis. One sampling scheme that is popular but challenging to implement is \emph{weighted random sampling without replacement}, also known as sampling with \emph{probability proportional to size} (PPS). There are well over 50 papers on this topic; see Hanif and Brewer~\cite{hb80} for a review of the older literature. In a fixed-size PPS sampling scheme, each item~$x_i$ is accompanied by an observable positive-valued auxiliary variable $w_i$, called the ``weight'' (or sometimes the ``size''), and the goal is to output a sample $S$ containing a specified number $n$ of items, i.e., a sample of size $|S|=n$, where each item in the population appears in the sample with probability proportional to its weight. An early motivation for PPS sampling was that use of fixed-size PPS samples when estimating a population total $\sum_{i=1}^N y_i$ via the Horvitz-Thompson formula leads to highly precise estimates when the weight of each item~$i$ approximates $y_i$ \citep[Sec.~3.6]{ssw92}. Recent work has indicated the usefulness of PPS samples for online management of supervised learning models~\citep{HentschelHT19}, where precise control over sample content is of paramount importance.

Unfortunately, it is not possible to simultaneously enforce both the sample-size requirement and the PPS property for all datasets. As a simple example, consider a dataset with six items $a_1,\ldots,a_6$ of weight 1 and six items $b_1,\ldots,b_6$ of weight 4, and suppose that we desire a sample of exactly 10 items. Denote by $p$ (resp., $q$) the appearance probability of an item $a_i$ (resp., $b_i$). Since the sample size $|S|$ equals 10 with probability~1, we have $\mean\bigl[|S|\bigr]=6p+6q=10$; if $q=4p$, then we must have $p=1/3$ and $q=4/3$, which is impossible since $q$ is a probability. If we try and fix this situation by choosing $q\le 1$ and setting $p=q/4$, so that the PPS property holds, then $\mean\bigl[|S|\bigr]\le 7.5<10$. 



Existing algorithms either assume the given weights are such that this trade-off does not occur, or they prioritize achieving the target sample size $n$ over enforcing the PPS property. Specifically, suppose that we have a universe of $N$ items with positive weights $w_{1}, \dots, w_{N}$. Systematic sampling and conditional Poisson sampling schemes assume the existence of $\pi_{i} = \rho\cdot w_{i}$ with $0 \leq \pi_{i} \leq 1$ and $\sum_{i=1}^{N} \pi_{i} = n$; see, e.g., \cite{hajek1964,Madow1949}. Under these assumptions both goals are simultaneously satisfiable, but the conditions do not allow for arbitrary values of the $w_i$. Almost all other approaches include all items if $N \leq n$ and otherwise implicitly or explicitly solve the equation $n = \sum_{i=1}^{N} \min(1, \tau \cdot w_i)$ for $\tau$ and include each item with probability $\min(1, \tau \cdot w_{i})$; see, e.g., \cite{Chao82, Chromy1979, CohenDKLT11, Rosen1997, Tille2019}. When $N \geq n$, this always creates a sample of size $n$, but can violate the PPS property to an arbitrary degree when item weights differ significantly from each other. For instance, in our previous example, $\tau = 2/3$ and thus items of weight~1 are $2/3$ as likely to be in the sample as items of weight~4, instead of $1/4$ as likely, as was desired. The typical result is that items with weights much higher than the average appear in the sample with probability~1, and that lower weight items are over-represented as compared to their higher-weight counterparts, as exemplified by items $a_1$--$a_6$, $b_1$--$b_6$ in our example (Somewhat confusingly, such methods are sometimes referred to in the literature as enforcing ``strict PPS''). \citet{Sunter77,Sunter86} also allows appearance probabilities for items having small weights to deviate from exact PPS.

We extend the set of PPS sampling schemes to allow for a different trade-off between controlling the sample size and controlling the sample contents. Our new method, \emph{Exact and Bounded PPS} (EB-PPS), strictly enforces the PPS property at all times while ensuring that the sample size never exceeds the target value $n$. That is, $n$ is now an upper bound on the sample size rather than the exact sample size.\footnote{\citet{Sunter77,Sunter86} also considers the case where $n$ is an upper bound on the sample size, but allows over-representation of heavy items.} EB-PPS ensures that each item~$x_i$ has appearance probability $\rho\cdot w_i$ as desired, where $\rho =\min(1/\max_i w_i,n/\sum_i w_i)$. If $\rho = n/\sum_i w_i$ then both a sample size of $n$ and the PPS property are simultaneously feasible, and our scheme yields the same appearance probabilities as those given above. If $\rho = 1/\max_i w_i$, then we prove that EB-PPS produces a PPS sample whose expected size is maximal and whose sample size variability is minimal over all possible PPS samples. Note that the latter situation occurs when $w_{max}/\bar{w}\ge N/n$, where $w_{max}$ and $\bar{w}$ are the maximum and average, respectively, of the $w_i$'s. The sample size therefore falls below the target $n$ in the presence of items having large relative weights. Note that, if the sample size becomes very small to the extent that it becomes problematic for the downstream application, then the small sample size serves as a signal to alert the user to the issue; for the other PPS methods, the composition of the sample will markedly diverge from the user's intent without any warning being given.

Thus our sampling scheme EB-PPS reverses the usual assumptions and treats proper data representation as more important than obtaining the maximal sample size. By bounding the sample size, EB-PPS, like prior schemes, helps control the time required for analytics over the sample and avoids storage overflows, especially when many samples are being maintained in parallel. However, the appearance probabilities are now ``as advertised'',  promoting user trust and allowing for easier reasoning in downstream applications (which can be much more complex than simply computing Horvitz-Thompson estimates).


In Section~\ref{sec:exClassifier}, we illustrate the advantages of EB-PPS sampling for a complex statistical-learning task. We show that, in general, PPS sampling can be used to train Bayes-optimal classifiers using standard techniques for 0-1 loss when the actual loss function is imbalanced. Prior sampling algorithms, which do not enforce the exact PPS property, yield classifiers that diverge from Bayes-optimality. As a result, EB-PPS yields better classification results. Moreover, EB-PPS achieves its superior results using sample sizes that are, on average, one third to one half as large as those produced by the other schemes. Thus a small, carefully curated sample can outperform a sample that is larger but less well designed.



EB-PPS has several important operational benefits that make it useful in practice. First, the algorithm works in the context of data streams: it views all items exactly once, does not need to know the dataset size in advance, and can forget all non-sampled items. For a fixed, finite data set, it follows that EB-PPS can produce a sample via a single pass through the data.
Second, the algorithm is efficient: we prove that the amortized processing time is $O(1)$ per item, which matches the optimal complexity of Chromy's algorithm~\cite{Chromy1979} for static datasets and 
improves upon the $O(\log \log n)$ time per item of the state-of-the-art VarOpt algorithm for streaming data~\cite{CohenDKLT11}. Finally, EB-PPS has low memory overhead and is simple to implement. The only data structure used throughout is an array of size $n$. So even when the weights are such that the Chromy or VarOpt algorithms would yield an exact PPS sample, there is a compelling argument to use EB-PPS instead.

The general approach embodied by EB-PPS originated in an effort to develop temporally biased sampling schemes with bounded sample size~\cite{HentschelHT19}. In that setting, the weight of an item initially equals 1 and decays over time. The current setting is more general in that the weights $w_1,w_2,\ldots,w_N$ are not necessarily monotonically increasing, but simpler in that the weights do not change over time.

%% file: ExactMarginalWeightedSampling.tex
\section{Overview of EB-PPS Sampling}

The algorithm presented below works sequentially over data streams of unknown size, so consider a sequence of items $x_1,x_2,\ldots$ with corresponding positive \emph{weights} $w_1,w_2,\ldots$ and for $t\ge 1$ let $U_t = \{x_1, \ldots, x_t\}$ be the set of items scanned so far. For any $t\ge 1$ we want to be able to produce an exact bounded PPS sample $S_t$ from $U_t$.

The first goal of EB-PPS sampling is to ensure that the appearance probability of each item $x_i$ is proportional to $w_i$ at all times or, equivalently,
\begin{equation}\label{eq:relWeights}
\frac{\pr(x_i \in S_t)}{\pr(x_j \in S_t)} = \frac{w_i}{w_j}
\end{equation}
for $t\ge 1$ and $i,j \leq t$. The other goal is to ensure that at each step~$t$ the sample size $|S_t|$ never exceeds $n$; the sample size should equal $n$ if feasible or, if not, then the sample size should have both maximal expected value and minimal variance relative to all possible bounded PPS samples.

As we have seen, rigorous enforcement of the PPS property can conflict with the goal of controlling the sample size. We therefore ``relax'' the problem by effectively allowing the sample size to take on fractional values; we then use a randomized procedure to deliver an integer-sized sample to the user. In more detail, EB-PPS maintains a data structure $L$ called a ``latent sample,'' from which we can extract an actual sample $S$ on demand. Formally, given a set $U$ of items, a \emph{latent sample} of $U$ having real-valued \emph{latent size} $C\ge 0$ is a triple $L=(A,\pi,C)$, where $A\subseteq U$ is a set of $\floor{C}$ \emph{full} items and $\pi\subseteq U$ is a (possibly empty) set containing at most one \emph{partial} item; $\pi$ is nonempty if and only if $C>\floor{C}$. In the following, we denote by $S,S',S_t$ samples extracted from $L,L',L_t$, by $(A,\pi,C)$, $(A',\pi',C')$, $(A_t,\pi_t,C_t)$ the components of $L,L',L_t$, and so on. We slightly abuse notation and write $x\in L$ for $L=(A,\pi,C)$ if $x\in A\cup\pi$. We similarly say that latent samples $L$ and $L'$ are \emph{disjoint} if the sets $A\cup\pi$ and $A'\cup\pi'$ are disjoint.

Latent samples are described in detail in the subsequent section. Here we cover the functionality of their three main methods:
\begin{enumerate}
    
    \item \textsc{Downsample}: Given a latent sample $L$ having latent size $C$ and a real number $\theta \in [0,1]$, the function \textsc{Downsample}($L,\theta$) produces a new latent sample $L'$ having latent size $C'=\theta \cdot C$ and satisfying
    \begin{equation}\label{eq:downsampProb}
      \pr(x \in S') = \theta\cdot\pr(x \in S)
    \end{equation}
    for every item $x$ in the population.
    \item \textsc{Union}: Given two disjoint latent samples $L'$ and $L''$, \textsc{Union}($L',L''$) produces a new latent sample $L$ such that
    \begin{itemize}
        \item $\pr(x \in S) = \pr(x \in S')$ for all $x \in L'$,
        \item $\pr(x \in S) = \pr(x \in S'')$ for all $x \in L''$, and
        \item $L$ has latent size $C=C' + C''$.
    \end{itemize}
    \item \textsc{Output}: For a latent sample $L$ having latent size $C$, the function \textsc{Output}($L$) produces a realized sample $S$ of expected size $C$ and of actual size either $\floor{C}$ or $\ceil{C}$. Thus if $C$ is an integer then $S$ is of exactly size $C$. 
\end{enumerate}
Latent samples are reminiscent of the PMR samples in \citep{Chromy1979}, but differ in that latent samples can decrease in size; this is why our scheme can support exact PPS sampling.

Algorithm~\ref{alg:EWbound} gives the EB-PPS sampling scheme. A sample can be materialized at any step $t$ by calling $\textsc{Output}(L)$ after execution of line~\ref{ln:unionLpTp}. 


\begin{algorithm}[ht]
\caption{\textsc{EB-PPS}$(n,\mathcal{S})$}\label{alg:EWbound}
\footnotesize{
$n$: Sample-size bound\\
$\mathcal{S}=\langle(x_1,w_1),(x_2,w_2),\ldots\,\rangle$: Stream of items and weights
\BlankLine
\BlankLine
Initialize: $L = L' = (\emptyset, \emptyset, 0)$,
\nonl$w_{\text{max}}=-\infty$, $W = 0$ \\
\For{$t\gets1,2,\ldots$}{
\tcc{compute new proportionality constant}
$w'_{\text{max}} \gets \max(w_{t}, w_{\text{max}})$\\
$W' \gets W + w_{t}$\\
$\rho' = \min(1/w'_{\text{max}}, n/W')$ \label{ln:setRhop}\\
\tcc{downsample old items and new item, union result}
\lIf{$W>0$}{$L' \gets \textsc{Downsample}(L,\rho'/\rho)$} \label{ln:downsampEWS1}
$T \gets (\{x_{t}\}, \emptyset, 1)$\label{ln:setT}\\
$T' \gets \textsc{Downsample}(T, \rho' \cdot w_t)$ \label{ln:downsampEWS2}\\
$L \gets \textsc{Union}(L', T')$ \label{ln:endRound} \label{ln:unionLpTp}\\
$(W, w_{\text{max}},\rho) \gets (W', w'_{\text{max}}, \rho')$ \label{ln:last} 
}
}
\end{algorithm}

The following result establishes the PPS property of the algorithm, as well as the sample size bound.

\begin{theorem}\label{th:exWeights}
For all $t\ge 1$ and $x_i$ with $1 \leq i \leq t$, we have $\pr(x_{i} \in S_{t}) = \rho_{t}\cdot w_{i}$, where
\[
\rho_{t} = \min\bigl(\frac{1}{\max_{1 \leq i \leq t} w_i}, \frac{n}{\sum_{i=1}^{t} w_i}\bigr).
\] Moreover, $|S_t|\le n$ for $t\ge 1$. 
\end{theorem} 
\begin{proof}
The proof is by induction. For $t=1$, the algorithm sets $\rho'=\rho_1=1/w_1$ in Line~\ref{ln:setRhop}. The unique sample $S'$ extracted from the latent sample $T$ defined in line~\ref{ln:setT} satisfies $\pr(x_1\in S')=1$ and the downsampling operation in Line~\ref{ln:downsampEWS2} then yields $\pr(x_1\in S'')=\rho_1\cdot w_1\cdot\pr(x_1\in S')=1$ by \eqref{eq:downsampProb}, where $S''$ is a sample extracted from $T'$. Note that $L'=(\emptyset,\emptyset,0)$ because $L'$ is initialized to this value and the downsampling operation in Line~\ref{ln:downsampEWS1} is not executed. It follows from the properties of the \textsc{Union} function that $\pr(x_1\in S_1)=\pr(x_1\in S'')$ and thus item $x_1$ is included in $S_1$ with probability $\rho_1\cdot w_1=1$. 

For $t>1$, we have that $\rho'=\rho_t$ from Line~\ref{ln:setRhop}. Thus for $i<t$ we have, after executing line~\ref{ln:downsampEWS1}, that $\pr(x_i\in S')=(\rho_t/\rho_{t-1})\cdot\pr(x_i\in S_{t-1})=(\rho_t/\rho_{t-1})\cdot\rho_{t-1}\cdot w_i=\rho_t\cdot w_i$ by \eqref{eq:downsampProb} and the induction hypothesis, where $S'$ is a sample extracted from $L'$. Similarly, for $i=t$, an argument similar to that given for $t=1$ shows that $\pr(x_t\in S'')=\rho_t \cdot w_t$, where $S''$ is a sample extracted from $T'$. The desired result then follows from the properties of the \textsc{Union} function. Note that the downsampling operation on Line~\ref{ln:downsampEWS1} is allowed since $\rho_t/\rho_{t-1}\le 1$ by positivity of the weights, and the downsampling operation on Line~\ref{ln:downsampEWS2} is allowed since $\rho_t \cdot w_t\le w_t/\max_{1\le i\le t}w_i\le 1$. This proves the first assertion of the theorem. To prove the second assertion, we note that, as discussed above, the first item $x_1$ is accepted into the latent sample with probability~1, the initial latent size is $C_1=1$, and $\rho_1=1/w_1$. By lines~\ref{ln:downsampEWS1}, \ref{ln:downsampEWS2}, and \ref{ln:unionLpTp} of Algorithm~\ref{alg:EWbound} and the properties of \textsc{Downsample}, it can be seen that the latent sizes obey the recursion $C_t=(\rho_t/\rho_{t-1})\cdot C_{t-1}+\rho_t\cdot w_t$. A simple inductive argument then shows that $C_t=\rho_t\cdot W_t$, where $W_t=\sum_{i=1}^t w_i$. Since $\rho_t\le n/W_t$ by definition, we have $C_t\le n$ for all $t\ge 1$. Finally, we have, by construction, that $|S_t|\le\ceil{C_t}\le n$ manifestly for $t\ge 1$.
\end{proof}

The following two theorems show that when EB-PPS sampling produces a sample of size less than $n$, the expected sample size is the maximum possible under the PPS constraint in \eqref{eq:relWeights} and the sample-size variance is the minimum possible given maximal expected size. 

\begin{theorem}\label{th:maxMean}
Let $H$ be any weighted sampling scheme that satisfies~\eqref{eq:relWeights} and denote by $S_t$ and $S^{H}_t$ the samples produced at step~$t$ by EB-PPS and by $H$. If $\mean\bigl[|S_t|\bigr] < n$, then $\mean\bigl[|S^H_t|\bigr]\leq\mean\bigl[|S_t|\bigr]$.
\end{theorem}

\begin{proof}
Since $H$ satisfies \eqref{eq:relWeights}, it follows that for each item $x_j$ with $j\le t$, the inclusion probability $\pr(x_j \in S^H_t)$ must be of the form $r_t\cdot w_{j}$ for some constant $r_t$ independent of $j$. Since $r_t\cdot w_j \leq 1$, it follows that $r_t\leq 1/\max_{1 \leq i \leq t} w_i$. The quantity on the right is exactly the constant Algorithm~\ref{alg:EWbound} uses for appearance probabilities when giving a sample of size less than $n$, and so the result follows.
\end{proof}

\begin{theorem}\label{th:minVar}
Let $H$ be any weighted sampling algorithm that satisfies~\eqref{eq:relWeights} and has maximal expected sample size $C_t < n$, and denote by $S_t$ and $S^{H}_t$ the samples produced at step~$t$ by EB-PPS and by H. Then $\var[|S^H_t|]\geq\var[|S_t|]$ for any $t\ge 1$.
\end{theorem}

\begin{proof}
Considering all possible distributions over the sample size having a mean value equal to $C_t$, it is straightforward to show that variance is minimized by concentrating all of the probability mass onto $\floor{C_t}$ and $\ceil{C_t}$. This is the sample-size distribution attained by EB-PPS.
\end{proof}

%% file: FractionalSamples.tex
\section{Operations on Latent Samples}

In this section we discuss the three methods \textsc{Downsample}, \textsc{Union}, and \textsc{Output}. We use the notation $\frc(C)=C - \floor{C}$ throughout. Proofs for all of the results in this section can be found in \cite{HentschelHT19}. 


\textsc{Downsample}: Given $\theta \in [0,1]$, the goal of downsampling $L=(A,\pi,C)$ by a factor of $\theta$ is to obtain a new latent sample $L'=(A',\pi',\theta\cdot C)$ such that, if we generate $S$ and $S'$ from $L$ and $L'$ via \textsc{Output}, the appearance probabilities are scaled down according to \eqref{eq:downsampProb}.
Theorem~\ref{th:downsamp} (later in this section) asserts that Algorithm~\ref{alg:downsamp} satisfies this property.

\begin{algorithm}[t]
\caption{\textsc{Downsample}$(L,\theta)$}\label{alg:downsamp}
\DontPrintSemicolon
\SetInd{0.5em}{1.5em}
{\footnotesize
$L=(A,\pi,C)$: input latent sample with $C>0$\;
$\theta$: scaling factor with $\theta \in [0,1]$
\BlankLine
\BlankLine
\lIf{$\theta=1$}{\Return $L'=(A,\pi,C)$}\label{ln:noDown}
$V\gets \textsc{Uniform}(0,1)$; $C' = \theta\cdot C$\;
\uIf(\Comment*[f]{no full items retained}){$\floor{C'}=0$\label{ln:no_full}}{
    \If{$V>\frc(C)/C$}{$(A',\pi')\gets\textsc{Swap1}(A,\pi)$}\label{ln:swap0}
    $A'\gets\emptyset$\;\label{ln:killA}
}
\uElseIf(\Comment*[f]{no items deleted}){$0<\floor{C'}=\floor{C}$\label{ln:no_delete}}{
    \If{$V>\bigl(1-\theta\cdot\frc(C)\bigr)/\bigl(1-\frc(C')\bigr)$\label{ln:no_swap}}{
        $(A',\pi') \gets\textsc{Swap1}(A,\pi)$\;\label{ln:convert}
        }
}
\Else(\Comment*[f]{items deleted: $0<\floor{C'}<\floor{C}$}){
    \eIf{$V\le \theta\cdot\frc(C)$\label{ln:normal}}{
        $A' \gets \textsc{Sample}(A,\floor{C'})$\;\label{ln:swapA}
        $(A',\pi')\gets\textsc{Swap1}(A',\pi)$\;\label{ln:swapB}
        }{
        $A' \gets \textsc{Sample}(A,\floor{C'}+1)$\;\label{ln:smpl}
        $(A',\pi')\gets\textsc{Move1}(A',\pi)$\;\label{ln:move}
        }      
}
\If(\Comment*[f]{no fractional item}){$C'=\floor{C'}$}{
    $\pi'\gets\emptyset$\;
    }
\Return $L'=(A',\pi',C')$
}
\end{algorithm}

In the pseudocode for Algorithm~\ref{alg:downsamp}, the \textsc{Uniform}$(a,b)$ function generates a random value uniformly from the interval $(a,b)$ and $\textsc{Sample}(A, n)$ samples $n$ items uniformly and without replacement from $A$. The subroutine $\textsc{Swap1}(A,\pi)$ moves a randomly selected item from $A$ to $\pi$ and moves the current item in $\pi$ to $A$. Similarly, $\textsc{Move1}(A,\pi)$ moves a randomly selected item from $A$ to $\pi$, replacing the current item in $\pi$ (if any). 

To gain some intuition for why the algorithm works, consider a simple special case: the goal is to form a latent sample $L' = (A',\pi', \theta\cdot C)$ from a latent sample $L=(A,\pi,C)$, where $C$ is an integer and $C'=\theta\cdot C$ is non-integer, so that $L'$ contains a partial item. In this case, we simply select an item at random (from $A$) to be the partial item in $L'$ and then select $\floor{C'}$ of the remaining $C-1$ items at random to be the full items in $L'$. By symmetry, each item $i\in L$ is equally likely to be included in $S'$, so that the inclusion probabilities for the items in $L$ are all scaled down by the same fraction, as required for \eqref{eq:downsampProb}. This scenario corresponds to lines~\ref{ln:smpl} and \ref{ln:move} in the algorithm, where we carry out the above selections by randomly sampling $\floor{C'}+1$ items from $A$ to form $A'$ and then choosing a random item in $A'$ as the partial item by moving it to $\pi$.

In the case where $L$ contains a partial item~$x^*$ that appears in $S$ with probability $\frc(C)$, the algorithm handles $x^*$ first, thus reducing the remaining problem to the prior case. In particular, $x^*$ should appear in $S'$ with probability $p=\theta\cdot \pr(x^*\in S)=\theta\cdot \frc(C)$. Thus, with probability~$p$, lines~\ref{ln:swapA}--\ref{ln:swapB} retain $x^*$ and convert it to a full item so that it appears in $S'$. Otherwise, in lines~\ref{ln:smpl}--\ref{ln:move}, $x^*$ is removed from the sample when it is overwritten by a random item from $A'$. In both cases, a new partial item is again chosen from $A$ in a random manner to uniformly scale down the inclusion probabilities. Depending on whether $x^*$ is kept or not, the problem then reduces to choosing $\floor{C'}$ or $\floor{C'} + 1$ items and the uniformity of the selection preserves property \eqref{eq:downsampProb} for all items in $A$. 

The if-statements in lines~\ref{ln:no_full} and ~\ref{ln:no_delete} cover corner cases of the algorithm in which (i) $L'$ does not retain any full items from $L$ and (ii) no items are deleted from the latent sample, e.g., when $C=4.7$ and $C'=4.2$. These cases are handled similarly to the previous case but special care is taken either because the item in $\pi$ cannot become a full item or cannot be deleted. In case~(ii), for example, no items are deleted from the latent sample, and the partial item $x^*\in\pi$ either becomes full by being swapped into $A'$ or remains as the partial item in $L'$. Denoting by $\gamma$ the probability of \textit{not} swapping, we have $\pr(x^*\in S')=\gamma\cdot\frc(C')+(1-\gamma)\cdot 1$. On the other hand, Equation~\eqref{eq:downsampProb} implies that $\pr(x^*\in S')=\theta\cdot\frc(C)$. Equating these two expressions shows that $\gamma$ must equal the expression on the right side of the inequality on line~\ref{ln:no_swap}.


Across all cases, property \eqref{eq:downsampProb} holds, and so we have the following theorem. 
\begin{theorem}\label{th:downsamp}
For $\theta\in[0,1]$, let $L'=(A',\pi',\theta\cdot C)$ be the latent sample produced from a latent sample $L=(A,\pi,C)$ via Algorithm~\ref{alg:downsamp}, and let $S'$ and $S$ be samples produced from $L'$ and $L$ via \textsc{Output}. Then $\pr(x\in S')=\theta\cdot\pr(x\in S)$ for all $x\in L$.
\end{theorem}

\textsc{Union}: The pseudocode for the union operation is given as Algorithm~\ref{alg:union}. The idea is to add all full items to the combined latent sample. If there are partial items in $L'$ and $L''$, then depending on the values of $\frc(C')$ and $\frc(C'')$ we transform them into either a single partial item (lines~\ref{ln:partialStart}--\ref{ln:partialEnd}), a full item (lines~\ref{ln:fullStart}--\ref{ln:fullEnd}), or a full plus partial item (lines~\ref{ln:fullPlusPartialStart}--\ref{ln:fullPlusPartialEnd}). Such transformations are done in a manner that preserves the appearance probabilities. Theorem~\ref{th:union} formalizes the main result below. 

\begin{algorithm}[ht]
\caption{\textsc{Union}$(L',L'')$}\label{alg:union}
\DontPrintSemicolon
{\footnotesize
$L'=(A', \pi', C')$: fractional sample of size $C'$\\
$L''= (A'', \pi'', C'')$: fractional sample of size $C''$\\
\BlankLine
\BlankLine
$C \gets C' + C''$\;\label{ln:plusC}
$V\gets \textsc{Uniform(0,1)}$\;
\uIf{$\frc(C') = \frc(C'') = 0$}{
    $A \gets A' \cup A''$\;
    $\pi \gets \emptyset$
}
\uElseIf{$\frc(C') + \frc(C'') < 1$}{ \label{ln:partialStart}
	$A \gets A' \cup A''$ \;\label{ln:unionA1}
	\leIf{$V\le\frc(C')/\bigl(\frc(C') + \frc(C'')\bigr)$}{
		$\pi \gets \pi'$
	}{
		$\pi\gets \pi''$
	}
}\label{ln:partialEnd}
\uElseIf{$\frc(C') + \frc(C'') = 1$}{\label{ln:fullStart}
     $\pi\gets\emptyset$\;
     \leIf{$V\le\frc(C')$}{
          $A\gets A'\cup A''\cup\pi'$
     }{
          $A\gets A'\cup A''\cup\pi''$
     }
}\label{ln:fullEnd}
\Else(\Comment*[h]{$\frc(C') + \frc(C'') >1$}){\label{ln:fullPlusPartialStart}
	\eIf{$V\le\bigl(1 - \frc(C')\bigr)\bigm/\bigl[\bigl(1 - \frc(C')\bigr) + \bigl(1 - \frc(C'')\bigr)\bigr]$}{
		$\pi = \pi'$\;
		$A \gets A' \cup A'' \cup \pi''$ \;\label{ln:unionA2}
	}{
		$\pi = \pi''$\;
		$A \gets A' \cup A'' \cup \pi'$ \;\label{ln:unionA3}
	}	
}\label{ln:fullPlusPartialEnd}
\Return $L=(A,\pi, C)$
}
\end{algorithm}

\begin{theorem}\label{th:union}
Let  $L' = (A', \pi', C')$ and $L'' = (A'', \pi'', C'')$, be disjoint latent samples, and let $L=(A,\pi,C)$ be the latent sample produced from $L'$ and $L''$ by Algorithm~\ref{alg:union}.  Let  $S'$, $S''$, and $S$ be random samples generated from  $L'$, and $L''$, and $L$ via \textsc{Output}. Then
\begin{romlist}
\item $C=C'+C''=\mean[S]$;
\item $\pr(x\in S)=\pr(x\in S')$ for all $x \in L'$; and
\item $\pr(x \in S) = \pr(x \in S'')$ for all $x \in L''$.
\end{romlist}
\end{theorem}

\begin{algorithm}[ht]
\caption{\textsc{Output}$(L)$}\label{alg:output}
\DontPrintSemicolon
{\footnotesize
$L=(A, \pi, C)$: fractional sample of size $C$\\
\BlankLine
\BlankLine
$V\gets \textsc{Uniform(0,1)}$\;
\uIf{$V\le \frc(C)$}{
$S\gets A\cup\pi$
}
\Else{$S\gets A$}

\Return $S$
}
\end{algorithm}

\textsc{Output}: We use Algorithm~\ref{alg:output} to create the sample $S$ from the latent sample $L=(A, \pi, C)$. The algorithm includes all $\floor{C}$ items of $A$ with certainty. If $\pi=\emptyset$, then $\mean\bigl[|S|\bigr]=|S|=\floor{C}=C$. Otherwise $\pi=\{x^*\}$ for some partial item $x^*$, and the algorithm generates $V$ from a Uniform$(0,1)$ distribution and includes $x^*$ in $S$ if $V \le \frc(C)$. In this case,
$\mean\bigl[|S|\bigr] = \bigl(1-\frc(C)\bigr)\cdot\floor{C}+\frc(C)\cdot(\floor{C}+1)=\floor{C}+\frc(C)=C$.

%% file: AlgorithmRuntime.tex
\section{Algorithmic Runtime}

The runtime performance of EB-PPS can be analyzed in terms of the average or maximum cost to process a scanned item. We focus on the cost of maintaining the latent sample as items are scanned and do not explicitly include the $\Theta(n)$ cost of materializing samples for the user. We also assume that the latent sample can fit in memory and, for $i\ge 0$, that the contents of $L_i$ can be freely overwritten when computing $L_{i+1}$. 
Under these assumptions we have the following result for the average per-item processing cost. 

\begin{theorem}\label{th:algRuntime}
For any sequence of $t$ items, the runtime of EB-PPS sampling is $O(t)$ so that the amortized execution cost is $O(1)$ per item. 
\end{theorem}

\begin{proof}
First observe that the execution of the \textsc{Union} operator in line~\ref{ln:unionLpTp} is a constant-time operation that involves (potentially) adding the single element in $T'$ to the current latent sample $L'$. Similarly, all of the other steps in EB-PPS are constant time operations, except for the \textsc{Downsample} operation.

For \textsc{Downsample}, note that if, as in line~\ref{ln:downsampEWS2} of EB-PPS, a latent sample with $C=1$ is downsampled to a new latent size $C'\le 1$, then either the algorithm immediately returns the original latent sample in line~\ref{ln:noDown} or executes a single swap in line~\ref{ln:swap0}, so that the call to \textsc{Downsample} has an $O(1)$ cost. In general, the only steps in \textsc{Downsample} that are not $O(1)$ operations are the executions of the \textsc{Sample} operator in lines~\ref{ln:swapA} and \ref{ln:smpl}. To analyze these costs, denote by $d$ the number of elements of $A$ discarded during a call to $\textsc{Sample}(A,m)$. We can implement \textsc{Sample} by storing the elements of $A$ in an array and adapting the algorithm of Fisher and Yates as implemented by \citet{Durstenfeld64} for randomly shuffling an array in a single pass. In a sequence of steps, the algorithm decrements a pointer~$i$ from $\textsc{Length}(A)$ down to 2. At each step, a random index $j$ is uniformly selected from $\{1,2,\ldots,i\}$, and elements $A[i]$ and $A[j]$ are swapped. In our setting, we can stop the algorithm after $d$ steps and view the $d$ rightmost elements of $A$ as the elements to be discarded; these elements can then be overwritten in subsequent steps. Thus we require $d$ swaps to execute \textsc{Sample}, so that the overall cost of executing \textsc{Downsample} is $O(d)$.

Thus, in EB-PPS, the downsampling operation in line~\ref{ln:downsampEWS2} of Algorithm~\ref{alg:EWbound} has cost $O(1)$ as discussed above, and the cost incurred in line~\ref{ln:downsampEWS1} is $O(d_i)$, where $d_i$ is the number of elements discarded from the latent sample when processing item $x_i$. Thus the total cost of processing items $x_1,\ldots,x_t$ is $O(D_t)$, where $D_t=\sum_{i=1}^t d_i$. The number of elements discarded from the latent sample is bounded by the number of elements inserted into the latent sample. Since at most one element is inserted per item scanned, we have $O(D_t)=O(t)$, and the desired result follows.
\end{proof}

The maximum cost to process an item is $\Theta(n)$, and indeed some items can incur such relatively high execution costs. However, since producing an output sample also incurs a cost of $\Theta(n)$, this execution cost does not seem prohibitive. Moreover, the $\Theta(n)$ cost is only incurred when the sample reaches its maximum size of $n$ items and then loses many items due to a new heavy item, switching the constant $\rho_t$ from $n /\sum_{i=1}^{t} w_i$ to $1/ \max_{1 \leq i \leq t} w_i$. When this scenario occurs, the sample is no longer capable of producing a sample of size $n$ without violating \eqref{eq:relWeights}, and algorithms other than EB-PPS would have $\Theta(n)$ over-represented items. 

%% file: ExampleClassifier.tex
\section{Example Use Case: Bayes-Optimal Classifiers for Imbalanced Loss Functions}
\label{sec:exClassifier}

The key difference between EB-PPS and prior sampling schemes is that EB-PPS prioritizes enforcing the PPS property over maintaining a fixed sample size. Which of these two goals to prioritize depends on the application. For simple problems such as Horvitz-Thompson estimation,
it is possible to include items to fill up the sample while correcting for deviations from the exact PPS property so that the estimator is still unbiased; the inclusion of more items then strictly lowers the variance of the estimator. For more complex applications, however, there is no easy way to correct for deviations from the exact PPS property, and including more items can actually hurt overall accuracy.

We now illustrate this latter scenario for a specific complex application: using samples to train non-parametric classifiers---e.g., k-Nearest Neighbor (kNN) or random forest (RF) classifiers---at scale and in the presence of imbalanced loss functions. In general, sampling is used to control training or inference time, or to rapidly deal with concept drift or class imbalances as in~\cite{HentschelHT18,HentschelHT19}. For the classifier application, sampling can also simplify model training under imbalanced loss functions. In more detail, recall that a classifier is a mapping $\mathcal{C}:\X\mapsto\Y$, where $\X$ is a set of observed \emph{features} and $\Y=\{1,\ldots,m\}$ is a set of \emph{classes}. Ground-truth data is generated by an (unknown) joint probability distribution $p$ over $\X\times\Y$. In the simplest and most highly studied case of 0-1 loss, a unit loss is incurred if an item is misclassified; otherwise the loss equals 0. In this setting, the best possible classifier---in the sense of minimizing the expected loss---is the \emph{Bayes-optimal classifier (BOC)} that, given an observation $x$, sets the predicted class $y$ as $y=\arg\max_i p(i\mid x)$; see, e.g., \cite[p.~21]{hastie2009elements}. That is, it is better to predict class $i$ than $j$ if and only if $p(i \mid x) / p(j\mid x)\geq 1$. Even for this simplest case, obtaining a BOC can be highly nontrivial; it is often approximated or obtained only in the limit as the number of training points grows.

Significant complications ensue when the losses are 
\emph{imbalanced}: for $i\in[1..m]$, misclassification of a class-$i$ item as having class~$j$ ($\not= i$) results in a loss $\ell_i$. The imbalance arises because certain misclassifications are more costly than others, such as when false negatives are more costly than false positives. Under this loss function, the BOC minimizes expected loss by predicting for each $x \in \X$ the class~$i$ with the highest value of $\ell_{i}\cdot p(i \mid x)$, so that it is better to predict class $i$ than $j$ if and only if $p(i \mid x) / p(j\mid x)\geq \ell_{j} / \ell_{i}$. PPS sampling is very useful in reducing the training problem under an imbalanced loss function to the simpler case of training under 0-1 loss. Given any type of classifier that converges to the BOC classifier under 0-1 loss, together with an imbalanced loss function $\ell=(\ell_1,\ldots,\ell_m)$, use of standard training methods for 0-1 loss over appropriate PPS samples results in convergence to the BOC under $\ell$. From a practical perspective, this significantly simplifies the training, because any intuition about generalization, overfitting, and choice of model for 0-1 loss now carries over to the setting of imbalanced loss. Specifically, if we choose the weight for each class~$i$ item to be $\ell_i$, PPS sampling induces a modified probability distribution $\yp$ over $\X\times\Y$ satisfying $\yp(i,x)\propto \ell_i\cdot p(i,x)$ so that
$\yp(i\mid x)/\yp(j\mid x)=(\ell_i/\ell_j)\cdot\bigl(p(i\mid x)/p(j\mid x)\bigr)$
for all $x\in\X$. Thus a BOC for 0-1 loss under data distribution $\yp$ translates to a BOC for the original imbalanced loss function $\ell$ under the original data distribution $p$.

For any sampling scheme that does not enforce exact PPS, either (i)~the resulting trained classifier is biased away from the Bayes-optimal decision and toward predicting classes having low misclassification losses, since they are overrepresented relative to their weights, or (ii)~a customized loss function or training procedure is needed. In principle, it might be possible to use more data while correcting for the over-inclusion of low-weight items, but this would significantly complicate classifier training, and for many types of classifiers (such as kNN and RF classifiers) it is quite unclear how to do so.

We demonstrate the advantage of EB-PPS sampling in the foregoing setting via a couple of simple numerical examples. We consider kNN and RF classifiers where the parameter $k$ for kNN is chosen as described below and all other kNN and RF parameters are set to the default values of their standard scikit-learn implementations~\cite{scikitLearnKNN,scikitLearnRF}.

\begin{table}[t!]
\parbox{.43\linewidth}{
\small
    \centering
    \begin{tabular}{rcccc}
    \toprule
         & \multicolumn{2}{c}{kNN} & \multicolumn{2}{c}{RF}\\
         & EB-PPS & VarOpt & EB-PPS & VarOpt\\
    \midrule
    Experiment 1 & \textbf{0.925} & 1.448 & \textbf{0.942} & 1.463\\
    Experiment 2 & \textbf{2.040} & 2.157 & \textbf{1.970} & 2.211\\
    \bottomrule
    \end{tabular}
    \caption{Average classifier loss over 1{,}000 trials. The standard error for all measurements is less than 0.01. }
    \label{tab:exp-results}
    }\quad\quad\quad
\parbox{.47\linewidth}{
\small
    \begin{tabular}{rcccc}
    \toprule
         & \multicolumn{2}{c}{kNN} & \multicolumn{2}{c}{RF}\\
         & EB-PPS & VarOpt & EB-PPS & VarOpt\\
    \midrule
    Sampling Time & \textbf{0.821} & 3.306 &  \textbf{0.827} & 3.336\\
    Train + Inference & \textbf{0.189} & 0.301 & \textbf{0.841} & 2.724\\
    \bottomrule
    \end{tabular}
    \caption{Mean runtime of 1{,}000 trials of Experiment 2 in seconds. The standard error for all measurements is less than 0.01. }
    \label{tab:exp-results2}
}
\end{table}

\paragraph{Experiment 1: Single-point prediction} We first conducted an experiment with $\Y=\{0,1\}$ and $\X=\Re^{5}$; our trained classifiers were used to predict $y$ for a single specified value of $x$. The experiment shows how deviating from specified inclusion probabilities can lead to suboptimal classifiers. The misclassification loss was set to $\ell_1=10$ for false negatives (1 misclassified as 0) and to $\ell_0=1$ for false positives (0 misclassified as 1). The ground-truth data distribution $p$ was defined as follows. We first generate an $x$ value as a sample from a normal distribution with mean 0 and covariance matrix $0.01I$ (where $I$ denotes the $5\times 5$ identity matrix) and then generate a $y$ value such that $p(y=1\mid x) = 0.15$ and $p(y=0\mid x)=0.85$, independent of $x$. (Thus any $x$ value will serve equally well as our specified test value.) We conducted 1000 trials for each type of classifier. For each trial, we first generated 100 training items as i.i.d.\ samples from $p$, then we created two training sets using VarOpt and EB-PPS sampling, respectively; for each sampling algorithm, we used weights equal to the loss values as discussed above and a maximum sample size of $n=50$. (Recall that, for VarOpt, items have inclusion probability $\min(1,\tau\cdot w_{i})$ where $\tau$ is such that $\sum_{i}^{N} \min(1,\tau\cdot w_{i}) = n$.) We then trained kNN and RF classifiers over the two samples (with $k=9$ for kNN) using standard techniques for 0-1 loss. Finally, we tested each trained classifier at $x=0$; note that an optimal classifier should predict $y=1$ as it incurs expected loss $0.85$ whereas predicting $y=0$ incurs expected loss $1.5$. We found that the majority of the 1000 EB-PPS-trained classifiers correctly predict $y=1$. The classification results are summarized in Table~\ref{tab:exp-results}. As can be seen, the average loss using EB-PPS sampling was roughly 36\% less than the loss with VarOpt sampling on both types of classifier. The reason
is simple: VarOpt over-represents class-0 items, which have low misclassification loss, and so many of the resulting classifiers predict $y=0$ even though this prediction is sub-optimal. In contrast, EB-PPS maintains exactly the desired inclusion probabilities so that most of the resulting classifiers correctly predict $y=1$. Thus, use of EB-PPS yielded classifiers superior to those produced via VarOpt, even though the average sample size was 23.5 (versus the full sample size of 50 for VarOpt). 

\paragraph{Experiment 2: Multi-point prediction} Our second experiment considers a more complicated setting where oversampling of low-weight items leads to a suboptimal classifier when the loss is averaged over multiple points in $\X$. We set $\Y=\{1,2,3\}$, $\X=\Re^9$, and $(\ell_1,\ell_2,\ell_3)=(100,10,1)$. The ground-truth distribution $p$ is as follows. For each class~$i$, we first create a Gaussian-mixture distribution $G_i$ with 10 equally-likely components $N(\mu_1,I),\ldots,N(\mu_{10},I)$, where the centroids $\mu_1,\ldots,\mu_{10}$ are chosen uniformly from $[0, 1]^9$ and $I$ is the $9\times 9$ identity matrix. To generate a data point $(x,y)$, we first pick a class $y$ from $\Y=\{1,2,3\}$ with probabilities $\{1/73,8/73,64/73\}$, respectively, and then generate $x$ as a sample from $G_y$. This setup was deliberately chosen to make the classification task very challenging, so that good samples are crucial to learning the the classification boundaries in $\X$. In each of 1000 trials, we generated 100,000 data points and then created two training sets using VarOpt and EB-PPS sampling with a maximum of $n=10{,}000$ data points in each sample. We again trained kNN and RF classifiers on the EB-PPS and VarOpt samples, but now $k$ was chosen to be the best-performing value from the set $\{1, 2, 3, 4, 5\}$ for EB-PPS and for VarOpt. We then tested the trained classifiers on 4,000 randomly selected points in $\X\times\Y$ and recorded the average loss per data point. These average per-point loss values were then averaged over the 1000 trials to compute an overall average loss per data point. For kNN, the minimal overall average loss for each sampling method was achieved for $k=2$; these losses are shown in Table~\ref{tab:exp-results} along with average losses for the RF classifier. As can be seen, the overall average loss is lower for EB-PPS than for VarOpt; this was true across all values of $k$. The average sample size for EB-PPS was 3343 versus a full sample size of 10{,}000 for VarOpt, which explains the shorter training-plus-inference times in Table~\ref{tab:exp-results2}. (We do not report runtimes for Experiment~1 because, being a stylized example, the training-set sizes are very small, so that runtime differences are not informative.)
Again, smaller samples can produce better results if well curated. Other choices of experimental parameter values, not reported here, resulted in similar results for both experiments.

%% file: conclusion.tex
\section{Conclusion}

We have provided a new weighted sampling scheme, EB-PPS, that prioritizes the PPS property over maintaining a fixed sample size, thereby expanding the set of known unequal-probability sampling schemes. The scheme enforces an upper bound on the sample size while keeping the sample size as large and as stable as possible.
We have shown the potential usefulness of our scheme in the setting of a complex classification problem.
In addition, EB-PPS is an easy-to-implement, one-pass streaming algorithm that
has the best known amortized execution cost per item. 
Thus, even when it is possible to
both maintain a specified sample size and enforce specified PPS item-inclusion 
probabilities,
the EB-PPS algorithm should be preferred over prior algorithms from a computational perspective.